\documentclass[a4paper,11pt]{article}
\usepackage{pos}
\usepackage{graphicx}
\usepackage{float}

\usepackage{color}

\newcommand*\dif{\mathop{}\!\mathrm{d}}

\title{Hierarchy in double SU(2) models}

\author*[a]{Clara \'Alvarez-Luna}
\author[a]{Jos\'e A. R. Cembranos}
\author[a,\dagger]{Juan Jos\'e Sanz-Cillero}

\affiliation{Departamento de F\'\i sica Te\'orica and IPARCOS, Facultad de Ciencias F\'\i sicas,\\
Universidad Complutense de Madrid, Ciudad Universitaria, 28040 Madrid, Spain}

\emailAdd{c.a.luna@ucm.es}
\emailAdd{cembra@ucm.es}
\emailAdd{jjsanzcillero@ucm.es}

\abstract{
In this work, we provide a simple model that studies the probability to obtain a given hierarchy between two scales. In particular, we work in a theory with an $SU(2)_L\times SU(2)_H\times U(1)_X$ gauge symmetry and two scalar doublets. By the Coleman-Weinberg mechanism, the gauge bosons and scalars obtain different masses, corresponding to the light and heavy sectors. We analyze the mass ratio of these sectors in order to discuss the hierarchy between them, and we define a probability associated to this hierarchy. We analyze different cases in which one of the sectors is fixed or both of them have free parameters, and also study the effect of including an interaction between them. We conclude that the probability of obtaining very large hierarchies is (logarithmically) small but not negligible.
}

\FullConference{
  The European Physical Society Conference on High Energy Physics (EPS-HEP2021), \\
  26-30 July 2021\\
  Online conference, jointly organized by Universität Hamburg and the research center DESY
}

\note{We thank F. J. Llanes-Estrada and V.~Sanz for useful comments. This work was supported by Grants FPU16/06960 (MECD), the MICINN (Spain) projects PID2019-107394GB-I00 (AEI/FEDER, UE) and PID2019-108655GB-I00 (AEI), the EU STRONG-2020 project [grant no. 824093], and STMS Grant from COST Action CA16108.}

\begin{document}
\maketitle

\section{Introduction}

In these proceedings we want to give some insight on the issue about a model in which we include two sectors of particles: light ($L$) and heavy ($H$)~\cite{toymodels}. For it we will study a double $SU(2)$ model, including the possibility of an interaction between them by means of an extra $U(1)$. 
In addition, we want to explore the possibility of producing the mass scales of the model by quantum effects, an approach that has been pursued in different frameworks, with the Coleman-Weinberg (CW) effective potential mechanism~\cite{Coleman:1973jx}, one of the most popular ones. Both authors showed how a theory that is symmetric when looking at the interactions present in the tree level Lagrangian can develop Spontaneous Symmetry Breaking (SSB) when the radiative corrections are taken into account (we will consider one-loop corrections, but higher order contributions can also be studied~\cite{Buttazzo:2013uya}). Then, our proposal is to start from a scale-less Lagrangian and make use of the CW mechanism~\cite{Coleman:1973jx,Weinberg:1973am,Gildener:1976ih,Chataignier:2018kay} to obtain SSB and generate the masses of the model. 
We will establish the complete Lagrangian for a double $SU(2)$ with an additional $U(1)$ symmetry group, including all the mixing terms that will give place to the gauge boson masses. 
Finally, we will study the parameter space of this model and analyze how probable is to obtain models that provide large mass differences between both sectors, that is, large hierarchies. 
For different works about hierarchy and naturalness problems read \cite{Ellis:1986yg,Barbieri:1987fn,Ciafaloni:1996zh,Casas:2014eca,Weinberg:1988cp,Bardeen:1995kv}.

\section{The model: $SU(2)_L \times SU(2)_H \times U(1)_X$}

We will work with one of the simplest models that can provide two different scales from radiative corrections: we will assume a model with a gauge symmetry group $\mathcal{G} = SU(2)_L \times SU(2)_H \times U(1)_X$ containing two complex scalar doublets under $SU(2)_L$ and $SU(2)_H$, $\Phi$ and $\Theta$, respectively.
The Lagrangian corresponding to this model is the following:
\begin{equation}
    \mathcal{L}_0 \,=\, |D_\mu\Phi|^2 +|D_\mu\Theta|^2 - V_0(\Phi,\Theta)\,, 
\end{equation}
where the covariant derivative includes the terms that give place to the gauge boson masses:
\begin{equation}
    \mathrm{D}^\mu_{L,H}=\partial^\mu-\frac{i}{2}g_{L,H}\sigma_a W^{a\mu}_{L,H}-\frac{i}{2}g_XQ_{L,H}X^\mu\,. 
\end{equation}
For the scalar fields it will be useful to choose the following orientation: $\Phi^T=(0,\varphi)/\sqrt{2}$ and $\Theta^T=(0,\eta)/\sqrt{2}$. 
Finally we complete the Lagrangian with the tree level potential:
\begin{equation}
    V_0(\varphi, \eta) = \frac{1}{4!} \lambda_L\varphi^4 
    +\frac{1}{4!} \lambda_H\eta^4 
    +\frac{1}{4!} \lambda_{LH}\varphi^2\eta^2 \,.
\end{equation}
From these expressions we can have the particle content of the model: the two scalars ($\varphi$ and $\eta$) and 7 gauge bosons whose masses $m_j$ depend on the scalar backgrounds $\varphi$ and $\eta$, and the gauge couplings. 
On the $L$-$H$ decoupled limit ($g_X = 0$): $m_{W_{L,\, j}}=g_L \varphi/2 $ ($j=1,2,3$); $m_{W_{H,\, j}}=g_H \eta/2 $ ($j=1,2,3$); $m_{X}=0$. 
On the general case ($g_X \neq 0$), we will have more involved expressions: $W^\mu_{L,\, 1} $, $W^\mu_{L,\, 2} $, $W^\mu_{H,\, 1} $ and $W^\mu_{H,\, 2}$ have the same $m$ as with $g_X=0$; however there is a mixing between $W^\mu_{L,\, 3}$, $W^\mu_{H,\, 3}$ and $X^\mu$ resulting in $Z^\mu_L,\, Z^\mu_H$ and $\hat{\gamma}^\mu$. 
There is a gauge boson $\hat{\gamma}^\mu$ that is always massless, while $Z_L^\mu$ and $Z^\mu_H$ have masses that are combination of the three gauge couplings $g_{L,H,X}$~\cite{toymodels}. 
If gauge boson loops dominate over the $\varphi$ and $\eta$ ones, the 1-loop effective potential becomes~\cite{Coleman:1973jx}:  
\begin{equation}
    V(\varphi,\eta) = V_0(\varphi,\eta) + \frac{3}{64\pi^2} \sum_{j=1}^7 m_j^4\left[ \ln\left(\frac{m_j^2}{\mu^2}\right)-\frac{5}{6}\right]\, ,
\end{equation}
where the sum runs over all the gauge boson masses of the model and $\mu$ is the renormalization scale. In addition, we need to include the CW restriction $|\lambda_j| \, < \,  \epsilon_{CW}\, \cdot \,  g_j^2 $ (gauge boson loop dominance), and perturbativity $g_j^2 \, < \,  \epsilon_{g^2} \,\cdot\,  4\pi\,  \equiv\,  g^2_{max}$\footnote{Further details on the restrictions can be found in~\cite{toymodels}.}. 
We can consider two cases: the simplest one, in which we analyze the implications of the model without any mixing between the $L$ and $H$ sectors, that is $g_X = 0, \lambda_{LH} = 0$, and also the mixed case. For the second one, we will consider $\lambda_{LH} = 0$ and a $g_X \neq 0$ coupling, which we will take to be small in order to obtain analytical expressions. 
For these cases, we also notice that the hierarchy between the $L$ and $H$ sectors is essentially given by the values of the parameters of both sectors, ($g_L, \lambda_L, g_H, \lambda_H$), while the mixing couplings tune the final precise position of the potential minimum.

\section{Phenomenology}

Now that we have the masses of the particles for both sectors, we want to study if it is possible to have a large difference between them, that is, a large hierarchy. For it, we define the hierarchy between the $L$ and $H$ sectors as follows:
\begin{equation}
    \mathfrak{R} = \frac{m^2_{W_H}}{m^2_{W_L}} =\frac{g_H^2 \langle\eta\rangle^2}{g_L^2\langle\varphi\rangle^2}\ .
\end{equation}
Thus, we will study situations in which the $L$ particles have light masses while the $H$ particles are at a higher mass scale, that is $\mathfrak{R}\gg 1$. We will see that if $g_X = 0$ and $\lambda_{LH} = 0$ simple analytical expressions are obtained; if $g_X \neq 0, \lambda_{LH} = 0$, analytical expressions can be obtained through a perturbative expansion in which are more involved than in the previous case; if $\lambda_{LH} \neq 0$ analytical expressions cannot be extracted and a more involved numerical calculation is needed. Therefore, the latter is left for future works and will no longer be discussed in these proceedings.

In addition to which values in our parameter space imply certain hierarchies, we want to analyze what is the probability to obtain those hierarchies. In this way we can check whether a given large value of $\mathfrak{R}$ corresponds to a fine-tuned choice of parameters or,  on the contrary, it is quite probable to obtain large hierarchies for a wide region of the parameter space. The main idea to carry out this calculation is that, as we will see, for different hierarchies we obtain bounded regions in our parameter space (that is also bounded by different restrictions). Then, we need to study the area (or hypervolume) of each region in relation to the total one of the parameter space. This will give us a measure of the probability associated to each hierarchy. 

\begin{figure}
\begin{center}
    \includegraphics[width=0.5\textwidth]{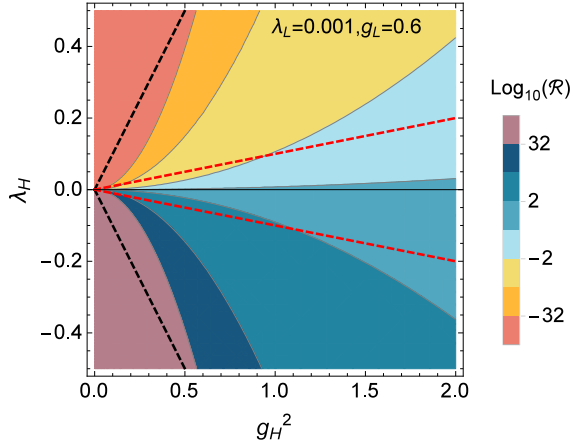}
    \caption{{\footnotesize Illustration of the allowed CW triangle in the $(g_H^2,\lambda_H)$ plane and the regions with a hierarchy $\mathfrak{R}\in[\mathfrak{R}_0,\mathfrak{R}_1]$. Here $g_{\rm max}^2=2$ and the dashed lines represent the CW restriction for $\epsilon_{\rm CW}=1$ (black) and $\epsilon_{\rm CW}=10^{-1}$ (red). 
    }}\label{fig:area}
\end{center}
\end{figure}

Our first approach consists on fixing the couplings of the $L$ sector instead of integrating the whole $L$–$H$ parameter space, corresponding to a situation where we have a certain fixed knowledge of the theory at low energies but still consider all possibly allowed configurations for the $H$ sector couplings. 
In the simplest case $g_X = 0$, we have an analytical expression for $\mathfrak{R}$:
\begin{equation}
    \mathfrak{R} = exp\left[\frac{128\pi^2}{27}\left(\frac{\lambda_L}{g_L^4}-\frac{\lambda_H}{g_H^4}\right)\right]\ 
    \mbox{, which implies} \quad  
    \lambda_H(g_H^2)\, =\, g_H^4\left(\frac{\lambda_L}{g_L^4} -\frac{27}{128\pi^2}\ln\mathfrak{R}\right)\, .
\label{eq:R}\end{equation}
This provides the constant hierarchy lines in the $(g^2_H, \lambda_H)$ plane that can be seen in Fig.~\ref{fig:area}. 
The described procedure can be summarized mathematically by a function 
$f(\mathfrak{R}_0,g_H^2;\epsilon, \alpha_L)$, which is the $\lambda_H (g_H^2)$ in~(\ref{eq:R}) restricted to the allowed CW region~\footnote{
This function is simply given by $f(\mathfrak{R}_0,g_H^2;\epsilon, \alpha_L)=\mbox{max}\{\lambda_H(g_H^2) + \epsilon_{\rm CW} g_H^2 \, , \,  0\}$ for $\lambda_H(g_{H}^2)\leq 0$, 
and $f(\mathfrak{R}_0,g_H^2;\epsilon, \alpha_L)=\mbox{min}\{\lambda_H(g_H^2) + \epsilon_{\rm CW} g_H^2 \, , \,  2 \epsilon_{\rm CW} g_H^2\}$ for $\lambda_H(g_{H}^2)\geq 0$. }. 
With this definition, by integrating this function up to the maximum value of $g_H^2$ we obtain the area associated to a certain hierarchy: 
\begin{equation}
    \mathcal{A}(\mathfrak{R}_0;\epsilon,\alpha_L) = \int\hspace*{-0.25cm}\int_{\mathfrak{R}\geq \mathfrak{R}_0} \dif g_H^2\, \dif \lambda_H =\int_0^{g^2_{max}} f(\mathfrak{R_0},g_H^2;\epsilon,\alpha_L)\dif g_H^2\ .
\end{equation}

The total area $\mathcal{A}_T(\epsilon)=\epsilon_{\rm CW}g_{\rm max}^4=16\pi^2 \epsilon_{\rm CW}\epsilon_{\rm g^2}^2$ is given by the integral of the region allowed by all the restrictions. 
With this, we define the conditional cumulative probability $\mathfrak{P}^{(\alpha_L)}_{cumul}$, for fixed $\alpha_L=(g_L^2,\lambda_L)$, as the ratio of the area with $\mathfrak{R} \in [\mathfrak{R}_0, \infty]$ and the total allowed area $\mathcal{A}_T(\epsilon)$ in the $(g^2_H, \lambda_H)$ plane (CW-triangle): 
\begin{equation}
    \mathfrak{P}^{(\alpha_L)}_{\rm cumul} \, =\, \frac{\mathcal{A}(\mathfrak{R}_0;\epsilon,\alpha_L)}{\mathcal{A}_T(\epsilon)} 
    \,=\, 
    \frac{1}{6}\left(\frac{27\ln\mathfrak{R}_0}{32\pi} -\frac{4\pi\lambda_L}{g_L^4}\right)^{-2}
    \,\stackrel{\mathfrak{R}_0\gg 1}{\simeq} \, 
    \frac{0.44}{\left(\log_{10}\mathfrak{R}_0\right)^2} \, ,
\label{eq:pcumul}\end{equation}
for large enough hierarchies, such that $\lambda_H(g_{\rm max}^2)< -\epsilon_{\rm CW} g_{\rm max}^2$. The integrals for smaller values of $\mathfrak{R}$ can also be easily computed. 
We emphasize that the dependence on the tolerance exactly cancels in this ratio for $\epsilon_{\rm CW}=\epsilon_{\rm g^2}=\epsilon$ (otherwise, a marginal dependence remains).  

If we include a mixing between both sectors through $g_X\neq0$ (with $Q_L,Q_H\neq0$), we will have some modifications of the previous expressions. In order to have analytical expressions we need to stay in the $g_X\ll1$ limit, that is, a small mixing; the validity of this limit has been also studied numerically but will not be discussed in these proceedings. In this case, for the hierarchy we have $\mathcal{O}(g_X^2)$ corrections for Eq.~(\ref{eq:R}) and Eq.~(\ref{eq:pcumul}) with a very similar result, that can be seen in \cite{toymodels}, with a complete plot showing the results for different choices of parameters.

\begin{figure}
\begin{center}
    \includegraphics[width=0.45\textwidth]{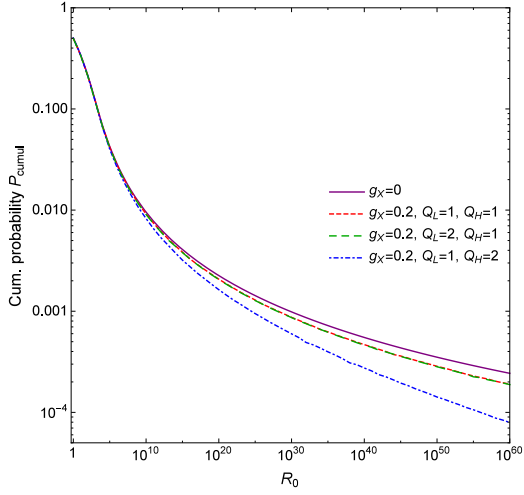}
    \caption{{\footnotesize Cumulative probability 
    for the integration of the whole $L$-$H$ hypervolume in the decoupled and coupled scenarios (both with $\lambda_{LH}=0$).}}\label{fig:cumul_prob_int}
\end{center}
\end{figure}

After the previous discussion on the $\alpha_L$–conditional probability, we conclude that the decoupled scenario ($g_X = 0$) seems to provide a fair enough approximation of the $L$-$H$ weakly interacting case. Now, we will study a more general case, in which both $L$ and $H$ sectors are integrated out. In a similar manner we derive the $\mathfrak{R}$ probability distribution from the integration to the whole $(g^2_L, \lambda_L, g^2_H, \lambda_H)$ allowed parameter space, by calculating the ratio of the hypervolume associated to a certain hierarchy compared to the total allowed hypervolume 
$\mathfrak{V}_T(\epsilon) = [\mathcal{A}_T(\epsilon)]^2$. 
For it, we need also to take into account that these parameters are also subject to the different restrictions that we have applied to the $H$ sector, i. e., the CW restriction and $g^2_{max}$, that made our model consistent. 
Notice that both the integrated volume and the total volume are proportional to $\epsilon^6$ (in the case with identical tolerances $\epsilon_{CW} = \epsilon_{g^2} \equiv \epsilon$), so the probability calculated as their ratio turns out to be tolerance independent. 
Starting from the previous definitions of $\mathcal{A}(\mathfrak{R}_0; \epsilon, \lambda_L,g_L^2)$ and $\mathcal{A}_T(\epsilon)$, we define the corresponding volumes, integrating on the $L$ parameter space:
\begin{equation}
    \mathfrak{V}(\mathfrak{R}_0;\epsilon) 
    \, =\,\int\hspace*{-0.25cm}\int\hspace*{-0.25cm}\int\hspace*{-0.25cm}\int_{\mathfrak{R}\geq \mathfrak{R}_0}  \ \dif{\lambda_L}\ \dif{g_L^2}\ \dif{\lambda_H}\ \dif{g_H^2} 
    \, =\,  
    \int_0^{g^2_{max}} \dif{g_L^2} 
    \int_{-\epsilon_{\rm CW}g_L^2}^{+\epsilon_{\rm CW}g_L^2} \dif{\lambda_L}\ 
    \mathcal{A}(\mathfrak{R}_0; \epsilon, \lambda_L,g_L^2)\ \ .
\end{equation}
Finally, the global cummulative probability for each hierarchy is defined by the ratio,  
\begin{equation}
    \mathfrak{P}_{\rm cumul} 
    \,=\,  \frac{\mathfrak{V}(\mathfrak{R}_0;\epsilon)}{\mathfrak{V}_T(\epsilon) }  
    \, \stackrel{\mathfrak{R}_0\gg 1}{\simeq }\,  \frac{1}{3}\left(\frac{32\pi}{27\ln\mathfrak{R}_0}\right)^2 \,\simeq \, \frac{0.87 }{(\log_{10}\mathfrak{R}_0)^2}  \, .
\label{eq:prob-cumul}\end{equation}
This $g_X=0$ result is plotted in in Fig.~\ref{fig:cumul_prob_int}. In this case, since the integration is more involved, we cannot obtain an analytical expression for the mixed case ($g_X\neq0$), however, the integration can be carried out numerically. Fig.~\ref{fig:cumul_prob_int} shows the outcome for different $Q_i$ choices.

Looking at the probability plots, we note that small hierarchies (near $\mathfrak{R} = 1$) are the most probable, and when we move away from it the probability decreases. Nevertheless, when we approach large hierarchies it decreases slowly enough to have non-negligible probabilities. We can understand this behavior better by looking at the analytical expression for $g_X=0$, where we can indeed see that it has an asymptotic logarithmic dependence.

\section{Conclusions}

In conclusion, we have studied a simple gauge model that allows large hierarchies between scales and constructed a probability that estimates how likely is to obtain a given hierarchy, defined by the hypervolume of the region of the parameter-space with that hierarchy, with the assumptions of the CW hypothesis and perturbativity \footnote{ 
Alternative probability studies of naturalness in supersymmetric frameworks can be found in \cite{Fichet:2012sn,Cabrera:2008tj,Ghilencea:2012qk}.}.
We have seen that wide regions of the parameter space give place to very different hierarchies, resulting on a probability of obtaining very large hierarchies that is suppressed, but only logarithmically, with  $\mathfrak{P}_{\rm cumul}\sim (\log_{10}\mathfrak{R})^{-2}$. This implies that, even though a small hierarchy between sectors is more likely, very large hierarchies cannot be excluded; e.g., scale differences such as the Planck scale over the electroweak scale, or even the Planck scale over the cosmological constant, would be only suppressed by probabilities $\mathfrak{P}_{\rm cumul}\sim 10^{-3}$--$10^{-4}$. Similar results would be obtained if different symmetry groups are included \cite{Fernandez:2015zsa}.



\begin{thebibliography}{99}

{\small 

\bibitem{toymodels}
C. \'Alvarez-Luna, J. A. R. Cembranos, and J. J. Sanz-Cillero, arXiv:2109.04955 [hep-ph].

\bibitem{Coleman:1973jx}
] S. R. Coleman and E. J. Weinberg, Phys. Rev. D7, 1888 (1973).

\bibitem{Buttazzo:2013uya}
D. Buttazzo, G. Degrassi, et al.
, JHEP 12, 089 (2013).

\bibitem{Weinberg:1973am}
E. J. Weinberg, Ph.D. thesis, Harvard U. (1973).

\bibitem{Gildener:1976ih}
E. Gildener and S. Weinberg, Phys. Rev. D13, 3333 (1976).

\bibitem{Chataignier:2018kay}
L. Chataignier, T. Prokopec, M. G. Schmidt, and B. \'Swie\.zewska, JHEP 08, 083 (2018).

\bibitem{Ellis:1986yg}
J. R. Ellis, K. Enqvist, D. V. Nanopoulos, and F. Zwirner, Mod. Phys. Lett. A 1, 57 (1986).

\bibitem{Barbieri:1987fn}
R. Barbieri and G. F. Giudice, Nucl. Phys. B 306, 63 (1988).

\bibitem{Ciafaloni:1996zh}
P. Ciafaloni and A. Strumia, Nucl. Phys. B 494, 41 (1997).

\bibitem{Casas:2014eca}
J. A. Casas, J. M. Moreno, S. Robles, K. Rolbiecki, and B. Zald\'ivar, JHEP 06, 070 (2015).

\bibitem{Weinberg:1988cp}
S. Weinberg, Rev. Mod. Phys. 61, 1 (1989).

\bibitem{Bardeen:1995kv}
W. A. Bardeen, in Ontake Summer Institute on Particle Physics (1995).

\bibitem{Fichet:2012sn}
S. Fichet, Phys. Rev. D 86, 125029 (2012).

\bibitem{Cabrera:2008tj}
M. E. Cabrera, J. A. Casas, and R. Ruiz de Austri, JHEP 03, 075 (2009).

\bibitem{Ghilencea:2012qk}
D. M. Ghilencea and G. G. Ross, Nucl. Phys. B 868, 65 (2013).

\bibitem{Fernandez:2015zsa}
G. Garc\'ia Fern\'andez, 
et al., Nucl. Phys. B 915, 262 (2017), [Err.: Nucl. Phys. B 949, 114755 (2019)].



}




\end{thebibliography}

\end{document}